\documentstyle[preprint,aps]{revtex}

\begin{document}
\draft
\title{Nuclear Magnetic Resonance Implemenations of Remote State Preparation of
Arbitary Longitudinal Qubit and Remote State Measurement of a Qubit }
\author{Xinhua Peng, Xiwen Zhu$^{\thanks{%
Corresponding author. {E-mail:xwzhu@wipm.ac.cn; Fax: 0086-27-87885291.}}}$,
Maili Liu, and Kelin Gao}
\address{Laboratory of Magnetic Resonance and Molecular Physics, Wuhan\\
Institute of Physics and Mathematics, The Chinese Academy of Sciences,\\
Wuhan, 430071, People's Republic of China}
\maketitle

\begin{abstract}
A qubit chosen from equatorial or polar great circles on a Bloch sphere can
be remotely prepared with an Einstain-Podolsky-Rosen (EPR) state shared and
a cbit communication. We generalize this protocal into an arbitrary
longitudinal qubit on the Bloch sphere in which the azimuthal angle $\phi $
can be an arbitrary value instead of only being zero. The generalized scheme
was experimentally realized using liquid-state nuclear magnetic resonance
(NMR) techniques. Also, we have experimentally demonstrated remote state
measurement (RSM) on an arbitary qubit proposed by Pati.
\end{abstract}


\section{Introduction}

Quantum entanglement is a novel attribute in quantum world that has no
classical counterpart. It is viewed not just as a tool for exposing the
weirdness of quantum mechanics, but as a potentially essential and valuable
resource for quantum communication. By exploiting entangled quantum sates,
one can performed tasks that are otherwise difficult or even impossible,
e.g., transmitted quantum states using local operations and classical
communication (LOCC) in absence of a quantum communications channel linking
the sender of the quantum state to the recipient. Among them are quantum
teleportation\cite{teleport} and remote state preparation (RSP)\cite
{remoArun,remoHoi,remoBennett}, which relies on the peculiar and apparently
non-local correlations inherent in entangled states.

Quantum teleportation\cite{teleport,teleportNMR,telExp1,telExp2} of an
unknown state is achieved from Alice to Bob by a shared EPR state (one ebit)
and two classical bits (cbits) communication, whereas remote state
preparation\cite{remoArun,remoHoi,remoBennett} of a quantum state known to
Alice but unknown to Bob can be completed by one ebit shared and only one
cbit communication. In previous literatures\cite
{remoArun,remoHoi,remoBennett}, the qubit in RSP can only be chosen from a
special ensemble, i.e., a qubit chosen from equatorial or polar great
circles on a Bloch sphere, which had been experimentally demonstrated by
liquid-state nuclear magnetic resonance techniques\cite{ourpaper_rsp}. While
RSP of an arbitrary qubit can succeed only half of the time. Though Bob
cannot entirely get an arbitrary qubit in remote state preparation, Pati\cite
{remoArun} also showed that it is possible for Bob to simulate with 100\%
efficiency any single-particle measurement on an arbitrary qubit known to
Alice but unknown to him, in virtue of an EPR pair and one cbit
communication.

In this paper, based on, we generalize Pati's scheme\cite{remoArun} for RSP
of a special ensemble of qubits into qubits from arbitrary longitudinal
lines on the Bloch sphere whose azimuthal angle $\phi $ can be an arbitrary
value instead of only being zero. And the generalized scheme of RSP has been
experimentally implemented via hydrogen and carbon nuclei in molecules of
carbon-13 labeled chloroform $^{13}$CHCl$_{3}$ over interatomic distances
using liquid-state nuclear magnetic resonance techniques. Moreover, we have
experimentally demonstrated remote state measurement on an arbitrary qubit
proposed by Pati\cite{remoArun}. These studies will be useful to understand
the foundations of quantum theory and perform effectively quantum
information processing with less resources.

\section{The Generalized Scheme of Remote State Preparation and Remote State
Measurement}

Let's begin with Pati's scheme of RSP\cite{remoArun} whose schematic diagram
is shown in Fig. 1a. The goal of RSP is for Alice to help Bob in a distant
laboratory to prepare a qubit state $|\psi \rangle $ known to her but
unknown to Bob with LOCC. An arbitrary single qubit can be represented by a
point on a Bloch sphere with two real parameters $\theta $ and $\phi $,
i.e., 
\begin{equation}
|\psi \rangle =\cos (\theta /2)|0\rangle +\sin (\theta /2)e^{i\phi
}|1\rangle ,
\end{equation}
where $\theta $ and $\phi $ are the polar and azimuthal angles on the Bolch
sphere, respectively. The scheme requires that Alice and Bob share an EPR
pair at first, 
\begin{equation}
|\psi ^{-}\rangle _{AB}=\frac{1}{\sqrt{2}}(|0\rangle _{A}|1\rangle
_{B}-|1\rangle _{A}|0\rangle _{B}),
\end{equation}
which can be rewritten in the qubit orthogonal basis $\{|\psi \rangle ,|\psi
_{\bot }\rangle \}$, 
\begin{equation}
|\psi ^{-}\rangle _{AB}=\frac{1}{\sqrt{2}}(|\psi \rangle _{A}|\psi _{\bot
}\rangle _{B}-|\psi _{\bot }\rangle _{A}|\psi \rangle _{B}),
\end{equation}
where $|\psi _{\bot }\rangle =\cos (\theta /2)|1\rangle -\sin (\theta
/2)e^{-i\phi }|0\rangle ,$ the complement qubit with $|\psi \rangle .$ Due
to her knowledge of $|\psi \rangle ,$ Alice can carry a single particle von
Neumann measurement by projecting onto the basis $\{|\psi \rangle ,|\psi
_{\bot }\rangle \}$ and then sends her measurement result to Bob. Depending
on the received classical information, Bob can recover the desired state $%
|\psi \rangle $ by deciding to do nothing $E$ or an operation $U$ which
converts $|\psi _{\bot }\rangle $ into $|\psi \rangle .$ Finally Bob's qubit
is prepared in the unknown state $|\psi \rangle .$ However, converting an
unknown $|\psi _{\bot }\rangle $ into $|\psi \rangle $ is an antiunitary
operation and thus RSP of an arbitrary qubit can succeed only half of the
time\cite{remoArun}. Fortunately, Pati\cite{remoArun} found a definite
unitary operator $U$ for a qubit of a special ensemble. For instance, for a
qubit chosen from equatorial $\left( \theta =\frac{\pi }{2}\right) $ and
polar great circles $\left( \phi =0\right) $ on a Bloch sphere, $U=\sigma
_{z}$ and $U=i\sigma _{y},$ respectively.

According to Pati's scheme, an preagreed convention about the prepared
qubit, such as the qubit chosen from equatorial or polar great circles on a
Bloch sphere, is made between Alice and Bob before performing RSP. If one
want to get a successful RSP\ all the time, the key is to find a universal
transformation $U$ to convert $|\psi _{\bot }\rangle $ into $|\psi \rangle .$
Then, does a universal transformation $U$ for the states expect for those
studied by Pati, exist?

Suppose that Alice and Bob, promise in advance that Alice would like to help
Bob remotely prepare a qubit chosen from the longitude with $\phi =\phi _{0}$
(an arbitrary value in the range of $[0,2\pi ))$ on the Bloch sphere, i.e., $%
|\psi \left( \phi _{0}\right) \rangle =\cos (\theta /2)|0\rangle +\sin
(\theta /2)e^{i\phi _{0}}|1\rangle .$ The entangled-pair sharing by both and
a von Neumann meausrement by Alice are the same as in the Pati's scheme.
Then Bob performs operations E or U according to the received one cbit
information from Alice after her measurement. For the qubit $|\psi \left(
\phi _{0}\right) \rangle $, we find a unitary operation 
\begin{equation}
U\left( \phi _{0}\right) =\left[ 
\begin{array}{ll}
0 & -e^{-i\phi _{0}} \\ 
e^{i\phi _{0}} & 0
\end{array}
\right] =i(\sin \phi _{0}\sigma _{x}-\cos \phi _{0}\sigma _{y})
\end{equation}
to correct $|\psi _{\bot }\left( \phi _{0}\right) \rangle $ into $|\psi
\left( \phi _{0}\right) \rangle ,$ that is, $|\psi \left( \phi _{0}\right)
\rangle =U\left( \phi _{0}\right) |\psi _{\bot }\left( \phi _{0}\right)
\rangle .$ Through the above procedure, RSP is successfully completed.
Therefore, a qubit chosen from any longitude can be remotely prepared by our
protocol. Two special cases can be dedured from Eq. (4), i.e.,
\begin{equation}
\phi =\{
\begin{array}{ll}
0, & U=i\sigma _{y}; \\ 
\frac{\pi }{2}, & U=i\sigma _{x}.
\end{array}
\end{equation}
It can be seen from Eq. (5) that the case with $\phi =0$ contained in our
generalized scheme is consistent with the Pati's scheme Our scheme includes
all points on the Bloch sphere.

The process of RSP\ stated above can succeed only for a preagreed ensemble
of qubits, however, Pati\cite{remoArun} showed that any single-particle
measurement on an arbitrary qubit can be remotely simulated with one ebit
and one cbit communication. This is attributed to Wigner's theorem\cite
{wigner} on symmetry transformations that shows the invariance of the
quantum-mechanical probabilities and transition probabilities under unitary
and antiunitary operations. Due to $\rho =\left| \psi \right\rangle
\left\langle \psi \right| =\frac{1}{2}\left( 1+\vec{n}\cdot \vec{\sigma}%
\right) $ and $\rho _{\bot }=\left| \psi _{\bot }\right\rangle \left\langle
\psi _{\bot }\right| =\frac{1}{2}\left( 1-\vec{n}\cdot \vec{\sigma}\right) $
for an arbitrary qubit, the probabilities of measurement outcomes in the
states $\rho $ and $\rho _{\bot }$ for an observable $\left( \vec{b}\cdot 
\vec{\sigma}\right) $ are given by 
\begin{equation}
\begin{array}{l}
P_{\pm }\left( \rho \right) =tr\left[ P_{\pm }\left( \vec{b}\right) \rho
\right] =\frac{1}{2}\left( 1\pm \vec{b}\cdot \vec{n}\right)  \\ 
P_{\pm }\left( \rho _{\bot }\right) =tr\left[ P_{\pm }\left( \vec{b}\right)
\rho _{\bot }\right] =\frac{1}{2}\left( 1\mp \vec{b}\cdot \vec{n}\right) 
\end{array}
,
\end{equation}
where $\vec{n}$ represents the direction vector, the Pauli operator $\vec{%
\sigma}=\left( \sigma _{x},\sigma _{y},\sigma _{z}\right) ,$ $\vec{b}$ is
the direction of measurement that Bob wants to choose and the projection
operator $P_{\pm }\left( \vec{b}\right) =\frac{1}{2}\left( 1\pm \vec{b}\cdot 
\vec{\sigma}\right) .$ From Eq. (6), Pati suggested the RSM scheme can be
always achieved by reversing the direction of $\vec{b}$ to make $P_{\pm
}\left( \rho _{\bot }\right) =P_{\pm }\left( \rho \right) $ when Bob gets
the state $\rho _{\bot }.$

\section{NMR Experiments\ and Results}

Our implementation of the generalized RSP and RSM\ schemes stated above were
performed using liquid-state NMR spectroscopy with carbon-13 labeled
chloroform $^{13}CHCl_{3}$ (Cambridge Isotope Laboratories, Inc.). The
spin-spin coupling constant $J$ between $^{13}C$ and $^{1}H$ is 214.95Hz.
The relaxation times were measured to be $T_{1}=4.8sec$ and $T_{2}=0.2sec$
for the proton, and $T_{1}=17.2sec$ and $T_{2}=0.35sec$ for carbon nuclei.
Hydrogen nucleus $(^{1}H)$ was chosen as the sender (Alice) and the carbon
nuclei $(^{13}C)$ as the receiver (Bob) in the experiments, transmitting the
state from $^{1}H$ to $^{13}C$. All experiments were performed on a
BrukerARX500 spectrometer with a probe tuned at 125.77MHz for $^{13}C$, and
at 500.13MHz for the $^{1}H$.

Using line-selective pulses and the gradient-pulse techniques\cite{xhpeng},
we initialized the quantum ensemble in an effective pure state (EPS) $\rho
_{0}$ from the thermal equilibrium. The shared EPR pair was experimentally
achieved by the NMR pulse sequence $\bar{Y}_{C}(\frac{\pi }{2})J_{CH}(\frac{%
\pi }{2})Y_{C}\left( \frac{\pi }{2}\right) \bar{X}_{C}\left( \frac{\pi }{2}%
\right) \bar{Y}_{H}\left( \pi \right) X_{H}\left( \frac{\pi }{2}\right) $ to
be applied from left to right, where $X_{H}(\frac{\pi }{2})$ and $\bar{X}%
_{H}(\frac{\pi }{2})$ denote $\frac{\pi }{2}$ and $-\frac{\pi }{2}$
rotations about $\hat{x}$ axis on spin $^{1}H$ and so forth, and $J_{CH}(%
\frac{\pi }{2})$ describes a time evolution of $1/2J_{CH}$ under the scalar
coupling between $C$ and $H$. We followed our approach detailed in Ref. \cite
{ourpaper_rsp} to the generalized RSP scheme. In the experiment, the single
particle von Neumann measurement in the qubit basis $\{|\psi \rangle ,|\psi
_{\bot }\rangle \}$ was simulated by using a two part procedure inspired by
Brassard et al.\cite{measure}. Using Alice's knowledge of the qubit, the
transformation $R_{H}^{+}\left( \phi _{0}\right) $ to rotate from the basis $%
\{|\psi \left( \phi _{0}\right) \rangle ,|\psi _{\bot }\left( \phi
_{0}\right) \rangle \}$ into the computational basis $\{|0\rangle ,|1\rangle
\},$ 
\begin{equation}
R_{H}^{+}\left( \phi _{0}\right) =\left( 
\begin{array}{cc}
\cos (\theta /2) & \sin (\theta /2)e^{-i\phi _{0}} \\ 
-\sin (\theta /2)e^{i\phi _{0}} & \cos (\theta /2)
\end{array}
\right) ,
\end{equation}
was realized by the NMR pulse sequence $X_{H}\left( \theta _{1}\right) \bar{Y%
}_{H}\left( \theta _{2}\right) X_{H}\left( \theta _{1}\right) $ with $\theta
_{1}=\tan ^{-1}\left( \tan (\theta /2)\sin \phi _{0}\right) ,$ and $\theta
_{2}=2\sin ^{-1}\left( \sin (\theta /2)\cos \phi _{0}\right) .$ The
projective measurement in the computational basis and the post-measurement
operation was equivalent with the conditional unitary operation $%
S=U_{C}\left( \phi _{0}\right) E_{+}^{H}+E_{-}^{H},$ where $E_{\pm }^{H}=%
\frac{1}{2}\left( 1_{2}\pm \sigma _{z}\right) $ and $1_{2}$ is a $2\times 2$
unit matrix, which was implemented by the pulse sequence $\bar{X}_{C}(\frac{%
\pi }{2})J_{CH}(\frac{\pi }{2})X_{C}\left( \frac{\pi }{2}-\phi _{0}\right) 
\bar{Y}_{C}\left( \frac{\pi }{2}\right) J_{CH}(\phi _{0})Y_{H}\left( \pi
\right) $ up to an irrelevant overall phase factor, where $J_{CH}(\phi _{0})$
describes a scalar coupling evolution of $\phi _{0}/\pi J_{AB}.$ The
measurement of the final state of spin $^{13}C$ is to record the spectra of $%
^{13}C$ nuclei and then integrate the entire multiplet after adjusting the
right phase\cite{Sharf}. In each experimental run for a given value of $\phi
_{0},$ a total of 13 qubit states with a $\theta $ increment of $\pi /12$
from $0$ to $\pi $ were studied. The experimental procedure was repeated for
different $\phi _{0}$ changed from $0$ to $2\pi $ with the increment of $\pi
/8.$ Thus RSP of a qibut chosen from different longitude on the Bloch sphere
were experimentally demonstrated. For each input state we recorded the total
NMR spectra of $^{13}C$ without and with a reading-out pulse $Y_{C}\left( 
\frac{\pi }{2}\right) ,$ and then deduced the real ($I_{x}$), imaginary ($%
I_{y}$) components of the former and the real component $(I_{z})$ of the
latter respectively. The measured data of the $I_{x},$ $I_{y}$ and $I_{z}$
components are plotted in Fig. 2, along with the theoretical expectations.
The state of $^{13}C$ nuclei is shown in fair agreement with the expected
form $\left| \psi \right\rangle ,$ which indicates the success of RSP.

For RSM, we performed the similar procedure only with the conditional
unitary operation $S$ omitted, i.e., measuring the NMR spectra of $^{13}C$
after the operation $R_{H}^{+}\left( \theta ,\phi \right) .$ Two multiplet
of $^{13}C$ can be labelled by two different subspaces of $^{1}H$\cite{Ernst}%
, the spectrum on the high frequency for the $\left| 0\right\rangle _{A}$
subspace and the one on the low frequency for $\left| 1\right\rangle _{A},$
and the corresponding states of $^{13}C$ after the operation $%
R_{H}^{+}\left( \theta ,\phi \right) $ are $\left| \psi _{\bot
}\right\rangle _{B}$ and $\left| \psi \right\rangle _{B},$ respectively. So
the outcomes in the state $\rho $ and $\rho _{\bot }$ can be obtained from
the measurements in different subspaces. In the experiments, we measured
three observables $\sigma _{x},$ $\sigma _{y}$ and $\sigma _{z},$ i.e., $%
\vec{b}=\left( 1,0,0\right) ,\left( 0,1,0\right) $ and $\left( 0,0,1\right) ,
$ respectively. As all NMR observables are traceless and the constant item
has no effect on the NMR signal, we got the results of RSM\ merely by
reversing the direction of the obtained $^{13}C$ signal in the $\left|
0\right\rangle _{A}$ subspace. We have studied a total of 221 points on the
Bloch sphere with a 13 by 17 rectangular grid of a $\theta $ spacing of $\pi
/12$ and a $\phi $ spacing of $\pi /8.$ Fig. 3 showed the experimental
results of RSM. A general observable $\left( \vec{b}\cdot \vec{\sigma}%
\right) $ is the linear sum of the three basic observables $\sigma _{x},$ $%
\sigma _{y}$ and $\sigma _{z}.$ In this experiments, Bob didn't use any
information about the original qubit $\left| \psi \right\rangle ,$ so RSM
was carried on an arbitrary qubit state.

As any point on the Bloch sphere can be expressed as the product form $%
I_{x}\sin \theta \cos \phi +I_{y}\sin \theta \sin \phi +I_{z}\cos \theta $
(apart from a constant unit matrix), the experimental results in Figs. 2 and
3 clearly shows the expected cosine and sine modulations, indicating that
the generalized RSP and RSM schemes are effective for all these input
states. However, it can be also seen that there are the discrepancies
between the experimental data and theoretical expectations, which are caused
by the static magnetic field and rf field inhomogeneities and the imperfect
calibrations of rf pulses, etc..

\section{Conclusion}

In summary, we have theoretically generalized Pati's scheme into an
arbitrary longitudinal qubit on the Bloch sphere with the azimuthal angle $%
\phi $ being an arbitrary value instead of only being zero and
experimentally implemented the generalized RSP protocol by using NMR quantum
logic gates and circuits in quantum information. Furthermore, RSM\ proposed
by Pati's were also experimentally investigated by NMR\ techniques, which
will be another powerful demonstration and a direct verification of Wigner's
theorem on symmetry transformations. Though some arguments about the
ensemble nature of NMR\ measurements and the short distances over which the
quantum information communication between spins (normally confined to
molecular dimensions in angstrom distance) happens, the concept and method
of quantum information transmission should be useful for quantum computation
and quantum communication.

\begin{center}
{\bf ACKNOWLEDGMENTS}
\end{center}

We thank A. K. Pati for bringing the topic of RSM to our attention and
Hanzheng Yuan, Zhi Ren, Daxiu Wei and Guoyun Bai for help in the course of
experiments.

\begin{center}
{\large Figure Captions}
\end{center}

Fig. 1 Schematic protocol for RSP proposed by Pati (a) and the corresponding
network of implementing RSP (b). In (a), the meter represents measurement,
and double lines coming out from it carry one classical bit $M$ (the
measurement outcome), while single lines denote qubits. In (b) $H$
represents the Hadamard gate and the conditional operation on a spin being
in the $\left| 1\right\rangle $ state and the $\left| 0\right\rangle $ state
are represented by a filled circle and an empty circle, respectively. $%
\oplus $ denotes the module 2 addition. $U$ and $R^{+}$ is denoted in the
text.

Fig. 2 Experimental results for RSP of qubits chosen from arbitary
longitudinal qubit.The left column represents the measured real (a) and
imaginary (b) parts of the normalized NMR\ signals from $^{13}C$ without any
reading-out pulse, and the real parts (c) of the same NMR signals after
applying a reading-out pulse $Y_{C}\left( \frac{\pi }{2}\right) ,$
respectively. The theoretical expectations are plotted in the right column.
All figures are shown as mesh and contour plots as a function of $\theta $
and $\phi $. The ordinate is in arbitrary units.

Fig. 3 Experimental data of remote state measurement for the observables (a) 
$I_{x}$, (b) $I_{y}$ and (c) $I_{z}$. The left and right columns represent
Bob's measurement outcome in the state $\rho $ and $\rho _{\bot }$ (by
reversing the direction of the observable). The theoretical expectations are
plotted in the middle column.

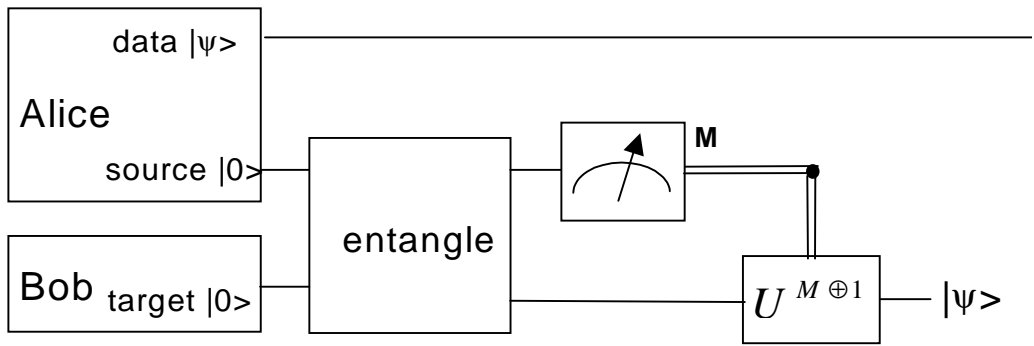
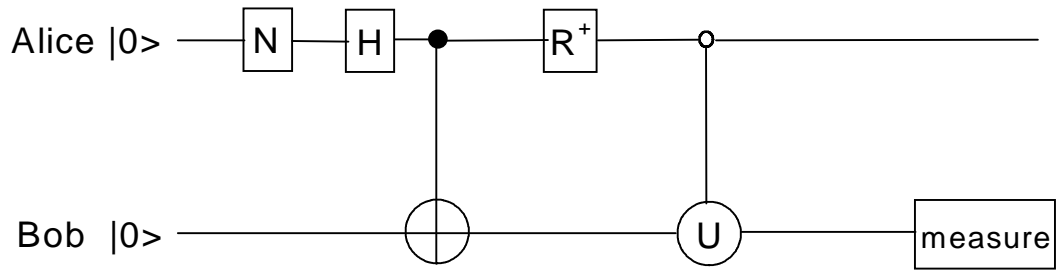
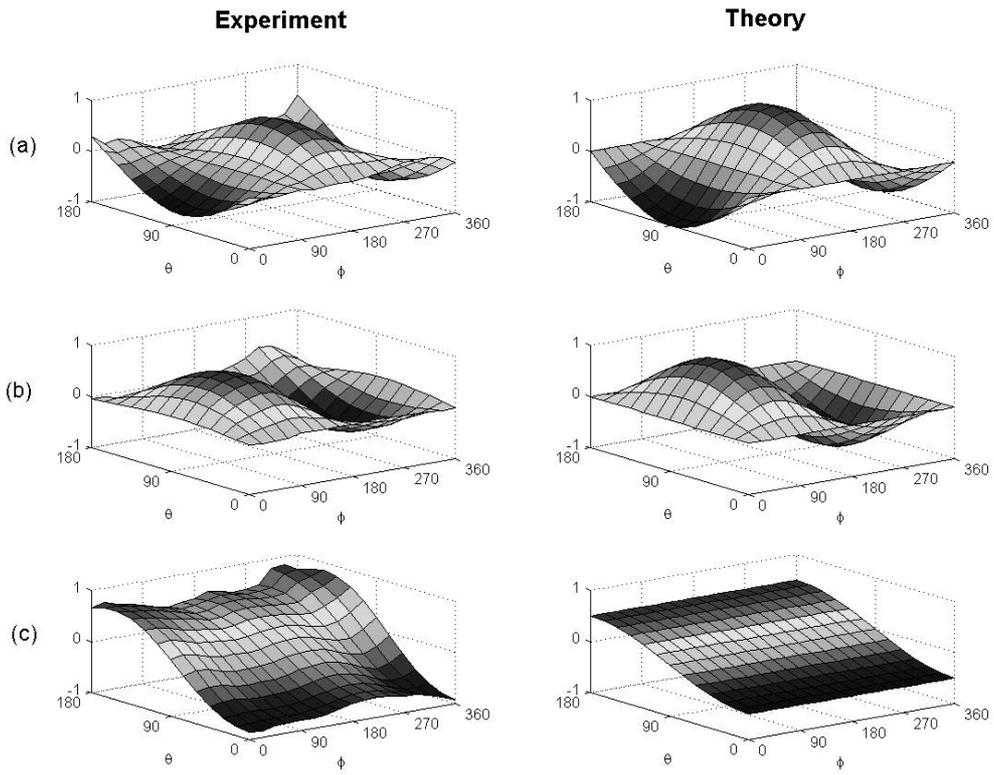
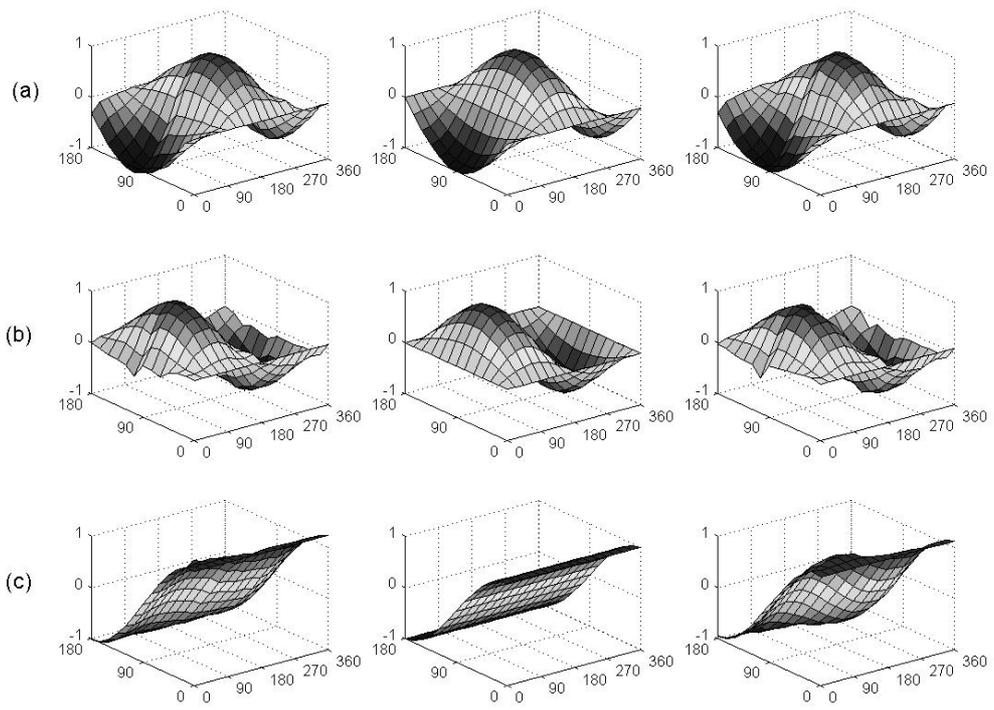
\end{document}